# AI-Based Speaking Assistant: Supporting Non-Native Speakers' Speaking in Real-Time Multilingual Communication

PEINUAN QIN, School of Computing, National University of Singapore, Singapore
ZICHENG ZHU, School of Computing, National University of Singapore, Singapore
NAOMI YAMASHITA, Social Informatics, Kyoto University, Japan
YITIAN YANG, School of Computing, National University of Singapore, Singapore
KEITA SUGA, Social Informatics, Kyoto University, Japan
YI-CHIEH LEE, Computer Science, National University of Singapore, Singapore

Non-native speakers (NNSs) often face speaking challenges in real-time multilingual communication, such as struggling to articulate their thoughts. To address this issue, we developed an AI-based speaking assistant (AISA) that provides speaking references for NNSs based on their input queries, task background, and conversation history. To explore NNSs' interaction with AISA and its impact on NNSs' speaking during real-time multilingual communication, we conducted a mixed-method study involving a within-subject experiment and follow-up interviews. In the experiment, two native speakers (NSs) and one NNS formed a team (31 teams in total) and completed two collaborative tasks—one with access to the AISA and one without. Overall, our study revealed four types of AISA input patterns among NNSs, each reflecting different levels of effort and language preferences. Although AISA did not improve NNSs' speaking competence, follow-up interviews revealed that it helped improve the logical flow and depth of their speech. Moreover, the additional multitasking introduced by AISA, such as entering and reviewing system output, potentially elevated NNSs' workload and anxiety. Based on these observations, we discuss the pros and cons of implementing tools to assist NNS in real-time multilingual communication and offer design recommendations.

## 1 Introduction

In international teams and collaborations, a single language is often adopted as the common medium of communication. However, for those participants who are non-native speakers (NNSs) of the language, engaging in multilingual communication can be particularly difficult, especially in terms of speaking. The main challenges in speaking stem from issues such as limited vocabulary, inadequate grammar, and poor language organization [5, 42, 52, 90, 95]. These linguistic barriers not only reduce NNSs' confidence in speaking but also increase their anxiety [1, 25, 35], significantly

Authors' Contact Information: Peinuan Qin, School of Computing, National University of Singapore, Singapore, e1322754@u.nus.edu; Zicheng Zhu, School of Computing, National University of Singapore, Singapore, zicheng@u.nus.edu; Naomi Yamashita, Social Informatics, Kyoto University, Japan; Yitian Yang, School of Computing, National University of Singapore, Singapore, yang.yitian@u.nus.edu; Keita Suga, Social Informatics, Kyoto University, Japan, suga.keita.77r@st.kyoto-u.ac.jp; Yi-Chieh Lee, Computer Science, National University of Singapore, Singapore, yclee@nus.edu.sg.

hindering their active participation in communication and, consequently, affecting the effectiveness of teamwork and collaboration [33, 58, 59, 87].

So far, a growing body of research has worked on supporting NNSs to speak, but it has not addressed the challenge of generating speaking content that NNSs face in such settings, as most existing efforts have focused on practicing speaking outside of real-time scenarios or increasing opportunities for NNSs to speak during real-time multilingual communication. For instance, some studies have helped NNSs improve their English speaking skills by enabling them to practice with advanced AI bots [38, 39, 72, 88]. Additionally, Li et al. developed a conversational agent to facilitate NNS participation in multilingual communication by automatically or manually opening the speaking floor when native speakers (NSs) have taken a certain number of consecutive speaking turns or upon NNSs' request [45]. While these studies provide valuable insights, the issue of NNSs struggling to express their ideas in the language used during real-time communication remains unresolved and urgently needs addressing.

Fortunately, recent advancements in AI technology, particularly in large language models (LLMs) such as the GPT series [70], accelerate AI-mediated communication (AIMC) [26] and open up significant opportunities for supporting NNSs' speaking. Specifically, AIMC involves an intelligent agent acting on behalf of a human communicator by altering, enhancing, or generating content to achieve communication objectives [26]. So far, AIMC has shown great potential in assisting NNSs with their writing content. For example, it can enhance the writing performance of NNSs by helping them achieve the appropriate tone in email writing through textual adjustments [12]. Given this effectiveness, we believe there is significant untapped potential for AIMC to support NNSs' speaking in real-time scenarios by helping with their speaking content, an area that remains largely unexplored.

Therefore, we developed an AI-based speaking assistant (AISA) that generates speaking references for NNSs in real-time communication by integrating NNS input queries, task background, and the conversation history. As this represents an initial attempt to support NNSs' speaking in real-time scenarios, we aim to explore how NNSs utilize the AISA, specifically focusing on identifying the input patterns of AISA, given that these interactions are central to NNSs' use of the tool. We also seek to understand how AISA influences NNSs' speaking in such dynamic settings through interviews and quantitative measures of speaking competence [22]. Notably, we excluded pronunciation from the speaking competence measure, as AISA's function is to provide speaking references and does not specifically focus on pronunciation. Additionally, given that NNSs' anxiety often stems from uncertainties about how to articulate [3, 75, 94, 95], and they often face a high workload due to multitasking in such settings, such as listening, comprehending, and speaking simultaneously [67], we are interested in exploring how introducing AISA influences their anxiety and workload.

To explore these questions, we conducted a mixed-methods study involving a within-subjects experiment and follow-up interviews. In the experiment, two NSs and one NNS formed a team (31 teams in total) and completed two collaborative tasks—one with access to the AISA and one without. During this process, we measured NNSs' experiences and usage of AISA through surveys and log data. Overall, our findings unveiled four distinct input patterns of AISA among NNSs during real-time multilingual communication—*seeking word translation, rationalizing decisions, stating viewpoints,* and using *only keywords*—each reflecting varying levels of effort and language preferences. Notably, the pattern *rationalizing decisions* posed a significant burden on NNSs, as demonstrated by their extended input durations and frequent modifications. Our findings also highlight the strategies employed by NNSs to integrate their native language with the language in use while utilizing AISA, which not only simplified input but also improved the logical structuring of complex ideas. Regarding NNSs' speaking, while quantitative data did not reveal a significant impact of AISA on NNSs' speaking competence, qualitative insights showed its contribution to the logical flow and richness of NNSs' speech. However, interviews also suggested that the additional multitasking demands introduced by AISA, such as inputting and reviewing system outputs, might potentially heighten NNSs' anxiety and workload.

In summary, this study contributes to the field of CSCW as follows:

(1) We designed an AI-based tool to support NNSs' speaking during real-time multilingual communication by providing speaking references for them. This represents a significant advancement over previous studies, which mainly focused on enhancing NNSs' speaking practice and increasing their speaking opportunities [39, 45, 72, 86, 88].
(2) We identified the input patterns of NNSs when using this AI-based tool, which offers design implications for crafting support tools tailored to these input patterns.
(3) We provided empirical support for the benefits of speaking content support offered by this tool and also the challenges of supporting NNSs in such real-time scenarios. These insights can aid in future developments aimed at supporting NNSs' speaking in real-time multilingual communication and broader AI-mediated speaking contexts.

## 2 Related Work

Speaking a common language can make it easier for people from different cultures to communicate, allowing for a more extensive dialogue and facilitating international multicultural teams [4]. However, people within a team or collaboration may exhibit varying proficiency levels in the common language. Particularly, non-native speakers (NNSs) of this language often face considerable obstacles due to their lower proficiency, especially in speaking, which can hinder their active participation and the overall effectiveness of the collaboration. In the following sections, we further unpack challenges NNSs encounter in speaking within multilingual communication, existing tools designed to support NNSs' speaking, and the potential of AI-mediated communication (AIMC) to enhance NNSs' speaking in real-time multilingual communication.

### 2.1 NNSs' Speaking Challenges in Multilingual Communication

NNSs frequently encounter significant speaking challenges in multilingual communication. The key challenges they face primarily stem from their insufficient vocabulary, inadequate grammar, poor language organization, and a lack of relevant cultural and knowledge background, all of which hinder their effective communication [8, 60, 85]. Specifically, insufficient vocabulary is identified as the most significant barrier, affecting an estimated 79.4% of language learners [42, 95]. Furthermore, difficulties in structuring thoughts into coherent sentences and the lack of grammatical proficiency add layers of complexity to these challenges [94]. Compounding these linguistic obstacles is the lack of familiarity with the relevant cultural contexts and subject matter knowledge, which exacerbates the difficulties NNSs face in articulating their thoughts effectively [85, 94].

Exacerbating the situation, such inadequate language skills further contribute to NNSs' speaking anxiety, lack of confidence, and hinder effective multilingual collaboration. Due to their limited speaking proficiency, NNSs often fear making mistakes or facing criticism [3]. For instance, an incomplete mastery of vocabulary can induce their fear of speaking, further reducing their confidence in communication [3]. Moreover, disparities in language proficiency within collaborative groups can lead to subgrouping or clustering [33], create unequal status and power dynamics [58, 59], and eventually erode trust among participants [87].

### 2.2 Tools to Support NNSs' Speaking

Recognizing these speaking challenges, a growing body of research has worked on supporting NNSs' speaking in multilingual settings, primarily focusing on enhancing their speaking practice [39, 72, 86, 88] and increasing their opportunities to speak during real-time communication [21, 45]. For example, Ruan et al. designed EnglishBot, an AI bot that practiced speaking with English learners by providing adaptive feedback [72]. It was found that users of EnglishBot exhibited significant improvements in fluency, meeting higher IELTS grading criteria [72]. Li et al. developed an agent designed to monitor conversation dynamics by tracking the speaking duration and frequencies of each participant. This agent intervenes when a certain participant, normally a native speaker (NS), dominates the discussion excessively. Furthermore, Duan et al. devised a feedback mechanism that notifies NSs when their speech becomes too complex or rapid, allowing NNSs more time to formulate and articulate their thoughts [21].

Despite the valuable insights provided by these studies, they fall short of addressing the core speaking challenges that NNSs face during real-time multilingual communication. Essentially, creating more speaking opportunities for NNSs does not necessarily equip them with the skills to express themselves. Even with extensive practice, real-time communication differs significantly from structured practice sessions, as NNSs often face significant pressure to respond to others promptly in real-time scenarios, striving to maintain the flow of dialogue to avoid prolonged silences and communication breakdowns [60]. Such immediacy of the interactions in real-time scenarios imposes a higher cognitive load on NNSs compared to practice, as NNSs need to simultaneously listen, understand, and organize their speech [21, 84]. Taken together, these underscore the urgent need for support that specifically targets and mitigates the speaking challenges NNSs encounter in real-time multilingual communication.

### 2.3 The Potential of AI-mediated Communication in Supporting NNSs' Speaking

The development of AIMC has shed light on potential solutions to the speaking challenges faced by NNSs during real-time multilingual communication. AIMC involves the use of intelligent agents that act on behalf of human communicators by altering, enhancing, or generating content to achieve specific communication objectives [26]. The recent advancement of AIMC is largely driven by breakthroughs in natural language processing (NLP), particularly the creation of advanced large language models (LLMs) [14, 63, 71]. By leveraging vast datasets and transformer architecture, LLMs have demonstrated superior performance in inference tasks, which enables LLMs to understand the nuances of ongoing conversations and provides users with contextually appropriate suggestions or modifications [50].

So far, AIMC has proven effective in providing support for written content through a wide range of functionalities, ranging from auto-correct, predictive text, grammar correction[1], to smart replies, auto-completion, and auto-responses [31, 41]. For instance, Fu et al. found that when individuals role-playing as staff members utilized higher-level AIMC supports—specifically, message-level suggestions rather than mere sentence-level ones—they achieved quicker response times in email replies and enhanced customer satisfaction [23]. By employing AIMC tools for brainstorming, interpreting the emotions behind received information, and expressing thoughts more accurately, users reported an increase in their confidence to communicate [24]. Moreover, AIMC has also been found to be beneficial in providing emotional support [77]. By enhancing the empathetic responses of support providers to those seeking support, these tools significantly improve the ability of individuals, especially those who initially struggle, to offer empathetic support [77].

In summary, the research mentioned above has demonstrated the capacity of AIMC to enhance communication content, although existing investigations of AIMC have primarily focused on written content. To harness the potential of AIMC in speaking, we developed an AI-based Speaking Assistant (AISA) to support NNSs' speaking in real-time multilingual communication.

[1]https://www.grammarly.com/

### 2.4 The Current Study

In light of the literature mentioned above, we designed an AISA in this study to provide speaking references for NNSs during real-time multilingual communication by integrating: (1) input queries from the NNS, (2) task background, and (3) the conversation history, leveraging the LLM from the OpenAI GPT-4 API.

Given that our study serves as the first attempt to provide speaking reference to NNSs in real-time multilingual communication, we conducted an exploratory study to investigate how NNSs utilize this supporting tool, particularly focusing on their input patterns, as interaction with AISA primarily occurs through input queries. Notably, despite our initial intent to offer speaking references for NNSs, they may also have their unique considerations, which could lead to input patterns that either align with or deviate from our expectations. Understanding these patterns is crucial as it enriches our comprehension of NNSs' speaking needs and will help improve the tool in a way that aligns with their actual usage. Consequently, we raise the following research question:

**RQ1: What are the input patterns of AISA among NNSs during real-time multilingual communication?**

Although AIMC has demonstrated positive effects in supporting text-based communication [23, 24, 31, 41, 77], its impact in real-time speaking scenarios remains unclear, where the difficulty of multitasking increases significantly [7, 43, 67, 83]. Therefore, it is crucial to thoroughly understand how AISA influences NNSs' speaking in such contexts. To this end, we formulate our second research question:

**RQ2: How does AISA affect NNSs' speaking during real-time multilingual communication?**

Additionally, we are interested in exploring how AISA affects NNSs' anxiety and workload when providing real-time assistance. As mentioned above, NNSs' speaking anxiety often stems from inadequacies in expressing themselves. By providing NNSs with speaking references, we seek to understand how such support influences NNSs' anxiety. Moreover, NNSs typically experience high workloads in real-time multilingual settings due to the need to listen, understand, and organize their speech simultaneously. Introducing a new tool may alter this dynamic, either alleviating or exacerbating their workload. To address these considerations, we formulate the third research question:

**RQ3: How does AISA influence NNSs' anxiety and workload during real-time multilingual communication?**

## 3 Methods

### 3.1 Overview

In this study, we conducted a mixed-methods study consisting of a within-subjects experiment and interviews. Participants were grouped into teams of two native English speakers and one non-native English speaker to reflect multilingual work environments where NSs often represent the majority, posing challenges for NNSs to navigate and find opportunities to speak up. Each team completed two survival tasks: one with the NNS using AISA and one without. We recorded AISA usage and team conversations. After each task, NNSs completed surveys. The experiment ended with interviews to further probe NNSs' experiences.

### 3.2 Participants

We recruited 31 English-NNS (Chinese native speakers) participants (19 females, 12 males) by posting an advertisement on online campus forums and 62 English-NS participants (34 females, 28 males) through local telegram communities. Overall, the NSs were all raised and educated in English-speaking countries, and these NNSs had all lived in an English-speaking country for less than one year. NNS participants reported their proficiency in English speaking with speaking scores in their IELTS test [2] (M = 6.175, SD = 0.294), ranging from 5.5 to 6.5. The mean age of the NS participants was 24.9 (SD = 4.0), and that of the NNS participants was 23.6 (SD = 1.9). Participants were compensated at a rate of $13 per hour. Before the study, we obtained approval from our Ethics Review Committee.

### 3.3 Experiment Task

Participants in each team engaged in three survival tasks: one practice task and two main tasks. These tasks, modified versions of the arctic, desert, and ocean survival tasks, were designed to simulate critical thinking and decision-making in a group collaboration [29, 44]. In these survival tasks, participants were presented with a hypothetical scenario where they had survived a catastrophic event, such as a plane crash, in an arctic, desert, or ocean environment. Participants were provided with a list of items, such as a compass and canvas, salvaged from the catastrophic event—four items in the practice task (Arctic Survival Task) and six in the main tasks (Desert and Ocean Survival Tasks). They were required to rank these items in order of importance for survival until rescue.

### 3.4 System Design

We conducted the experiment on our custom-built web interface, which allowed participants to view real-time transcripts and enabled NNSs to utilize the AISA during the practice task and one of the two main tasks.

*3.4.1 Web Interface.* To facilitate our within-subjects experiment, we developed two distinct web interfaces: one where NNSs can utilize AISA and another where AISA is not available. Fig. 1 illustrates the web interface with AISA. In this interface:

- (1) *Basic Information Header.* Displays information including the experimental room, the participant's username, user identifier, their role as NNS or NS, and the specific survival game (Arctic, Desert, or Ocean).
- (2) *Game Instruction Button.* The game instruction is accessible via a button that, when clicked, expands detailed instructions for the survival game on the right.
- (3) *Transcription Panel.* Displays the ongoing transcript, which indicates who says each statement.
- (4) *AISA Input Box.* Allows users to enter queries in various languages and formats.
- (5) *AISA Output Panel.* Displays the generated speaking reference.
- (6) *Items Panel.* Lists the survival items for easy access and discussion.

*3.4.2 Real-time Transcript.* As part of our web interface, we implemented a real-time transcription through the Automatic Speech Recognition (ASR) technology developed with AssemblyAI [3] (See Fig. 1 (3)). This use of real-time transcripts follows prior research [27, 64, 65, 78], which demonstrated that such technology could improve NNSs' comprehension of conversational context and enable NNSs to participate more effectively in multilingual discussions.

[2]The international standardized test of English language proficiency for non-native English language speakers, evaluated across four aspects: listening, writing, speaking, and reading, with a score range of 1-9 for each aspect.
[3]https://www.assemblyai.com/

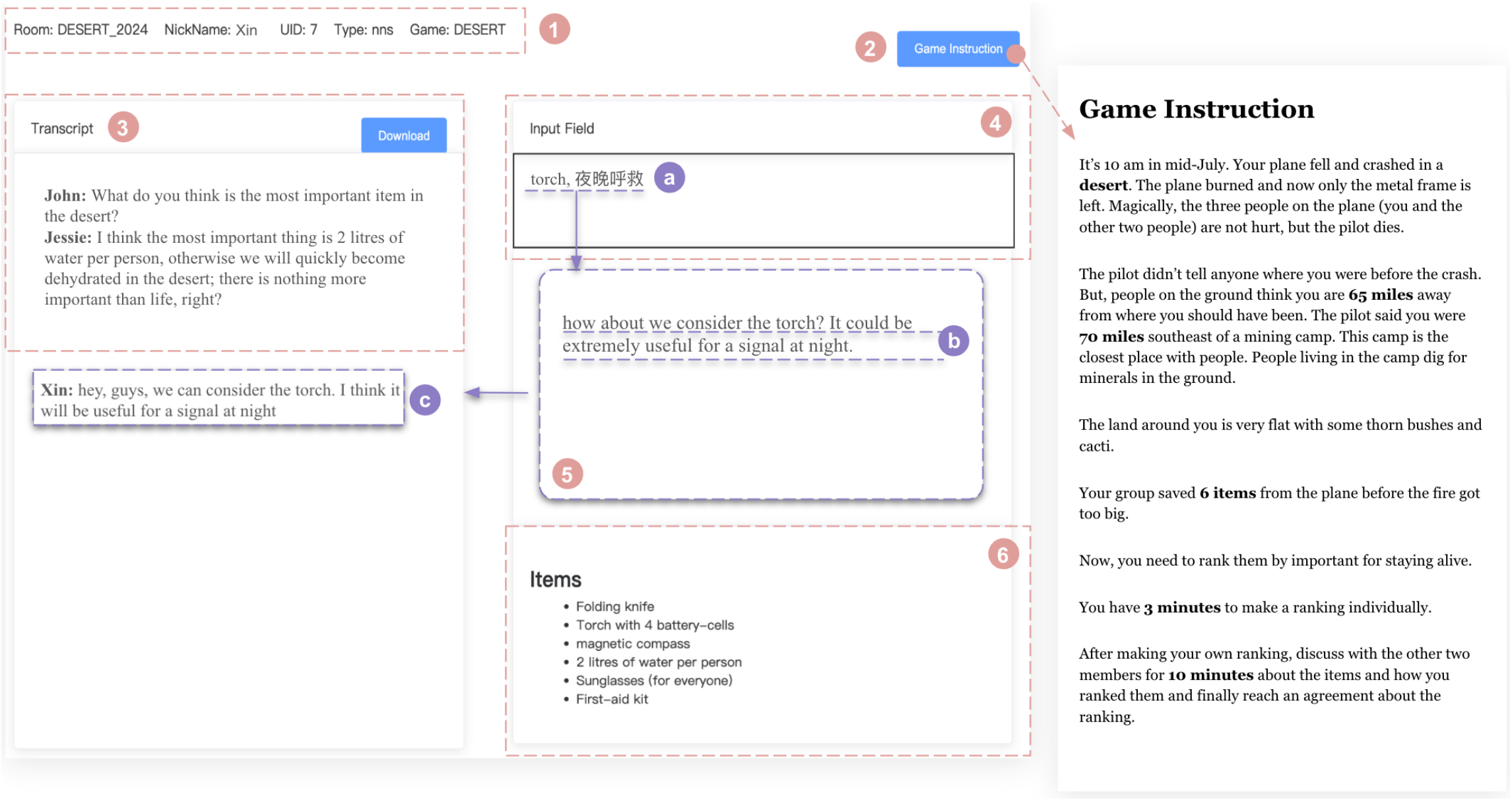

Fig. 1. Web interface with AISA. Where ① to ⑥ represents the components of the interface; ⓐ to ⓒ represents NNSs' operation steps when using AISA. Specifically, ① basic information header; ② game instruction button. When users click this button, detailed instructions for the survival game will be expanded; ③ transcript panel; ④ AISA input box; ⑤ AISA output panel; ⑥ items panel. Here, John and Jessie are NSs, while Xin is a NNS. As for the demonstration of NNSs' use of AISA, Xin displays three operation steps. ⓐ Xin inputs a query (The Chinese phrase "夜晚呼救" means "calling for help at night"). ⓑ Then Xin receives a speaking reference from AISA, and upon speaking, ⓒ the transcript automatically updates accordingly.

In our study, when participants talked, the ASR system converted their speech into text in real time and showed it on the transcript panel. The transcript included the username specified by the user to indicate who was speaking.

*3.4.3 AI-Based Speaking Assistant.* We designed AISA to provide speaking references for NNSs. The interface of AISA consists of two components: (1) an input box where NNSs can enter queries for which they need assistance (See Fig. 1 ④), and (2) an output panel that displays the generated speaking references (See Fig. 1 ⑤). The workflow of using AISA includes three steps: (See Fig. 1 ⓐ) NNSs input their queries into the input box, where they can enter keywords, phrases, or sentences in various languages. (See Fig. 1 ⓑ) Upon receiving these inputs, AISA generates speaking references mainly based on: (1) the inputs from the NNSs, (2) the task background, and (3) the conversation history. An example of a prompt transmitted to GPT-4 is shown in Table 1. The generated speaking references are then displayed in the output panel. Finally, (See Fig. 1 ⓒ) NNSs can choose whether and how to speak based on the generated reference and the transcript automatically updates.

*3.4.4 Prompt Design.* Our prompt consisted of four parts: (1) background, (2) conversation, (3) self-introduction, and (4) requirements.

Regarding the background and conversation, since AISA needed to generate context-based suggestions, it was essential to provide AISA with an understanding of the context, which was established through the "background" and "conversation" sections. Specifically, the "background" section contained task-related information. For example, in the desert survival game, we provided the following details:

**Background:**
[GAME INSTRUCTION]

**Self-introduction:**
I am a non-native speaker. My name is [NNS'S NAME].

**Requirement:**
I'm participating in the [SURVIVAL GAME NAME] and will be engaging in discussions. I will send you my conversations with other participants, which may include off-topic discussions. I will provide keywords or partial sentences representing the content I intend to communicate, and I want you to:

- Generate complete sentences in English based on my provided keywords and incomplete sentences, the context of the survival task, and the flow of the overall conversation. I will then use these sentences in my interactions with other participants.
- Do note that when generating these sentences, you're role-playing as [NNS'S NAME].
- Generate sentences that are more conversational and informal, reflecting the spoken language, not written formal language.

**Conversation:**
[TRANSCRIPT]

Table 1. **Prompt template transmitted to GPT-4 for generating speaking references.** The placeholders within brackets [ ] are dynamically replaced with real-time data during the experiment. Specifically, [GAME INSTRUCTION] is replaced with detailed survival game instructions tailored to the game context—ocean, desert, or arctic. [NNS'S NAME] is replaced with the specified username of the NNS to help the system accurately identify their speech within the transcript. [SURVIVAL GAME NAME] is updated with "ocean survival", "desert survival", or "arctic survival" accordingly. [TRANSCRIPT] is replaced by the actual real-time transcript of the ongoing discussion, where the NNS's input appeared at the end of it and was formatted as [USER NAME - Keywords]: [The input content of NNS].

> *I'm participating in a desert survival game. It's mid-July at 10 am, and my plane has crash-landed in a desert. Only the metal frame of the plane remains after a fire. Me and two others survive unscathed, but the pilot is dead. No one knew our location before the crash, but estimates suggest we're 65 miles off course. The pilot mentioned being 70 miles southeast of a mining camp, the closest inhabited location. The terrain is flat with occasional thorn bushes and cacti.*
>
> *My group managed to save six items from the plane before the fire spread. Now, our task is to rank the three most critical items for survival. Once I've made my list, I must discuss and reach a consensus on the item ranking with the other two survivors. The six items are: 1. Folding knife 2. Torch with 4 battery-cells 3. Magnetic compass 4. 2 litres of water per person 5. Sunglasses (for everyone) 6. First-aid kit*

The "conversation" section included the chat history between the NNS and NS. The conversation was formatted as follows: *[USER NAME]: [transcript content of this person]*, arranged in the actual speaking order. The NNS's input appeared at the end of the conversation and was formatted as: *[USER NAME - Keywords]: [The input content of NNS]*. An example of a conversation is provided below:

```
[John]: It's so hot out here. We definitely need water, or we won't last long.
[Jessie]: Agreed, but I'm also worried about the sun. Sunglasses might help us see better and avoid headaches.
[John]: That's true. But what about tools? A compass could be essential if we need to find our way out.
```

```
[Jessie]: Hmm, a compass is helpful, but if we don't have water, it won't matter much.
Maybe we prioritize water first?
[John]: Yeah, water first, then we can think about other items.
[John]: What do you think, Xin? Do you have any thoughts on what's most important?
[Xin - Keywords]: torch, 夜晚呼救
```

Regarding the self-introduction, the purpose was to help the LLM identify which parts of the historical chat record were spoken by the NNS and which by the NS. Therefore, we included the statement: "*I am a non-native speaker. My name is [NNS'S NAME].*"

Regarding the requirement, first, considering that NNS might provide keywords or fragmented sentences instead of complete sentences, we instructed the LLM to generate complete sentences based on the NNS's keywords and incomplete thoughts. The prompt specified: "*I will provide keywords or partial sentences representing the content I intend to communicate. Generate complete sentences in English based on my provided keywords and incomplete sentences, the context of the survival task, and the flow of the overall conversation. I will then use these sentences in my interactions with other participants.*" Second, during the pilot test, we observed that participants occasionally engaged in off-topic conversations, which affected the consistency of the LLM's generated content. For instance, some participants talked about scenarios and items related to their own experiences but irrelevant to this task. As a result, the LLM generated suggestions accordingly, deviating from the task background and items given. To address this, we added the instruction: "*I will send you my conversations with other participants, which may include off-topic discussions.*" Additionally, when generating responses based on the NNS's input, the LLM tended to adopt a default chatbot tone and perspective. To ensure the output reflected the NNS's first-person perspective, we instructed the LLM: "*Do note that when generating these sentences, you're role-playing as [NNS'S NAME].*" Moreover, since the conversation context was in English, we instructed the LLM to always output responses in English. Finally, because the setting involved spoken dialogue rather than formal writing, we requested the LLM to generate more conversational and informal sentences by including: "*Generate sentences that are more conversational and informal, reflecting spoken language, not written formal language.*"

We conducted multiple rounds of testing on the above prompt and confirmed that it achieved the intended effects.

*3.4.5 Example Use Scenario.* To illustrate the scenario in which NNSs utilize AISA, we present an example use scenario based on Fig. 1. This scenario involved three participants: John and Jessie, both NSs, and Xin, the NNS, participating in a desert survival game. During the game, John posed the question, *"What do you think is the most important item in the desert?"* Xin intended to suggest using a torch for signaling at night but struggled to express this in English. He inputted his idea into the system's input field as *"torch, 夜晚呼救"* using a mix of English and Chinese. The AISA then integrated the current conversation context along with Xin's input and generated a speaking reference from Xin's perspective: *"How about we consider the torch? It could be extremely useful for a signal at night."* Based on this reference, Xin stated, *"Hey guys, we can consider the torch. I think it will be useful for a signal at night."*

## 3.5 Procedure

In this study, we began by introducing the tasks and guiding participants through the consent form via Zoom. Participants were then divided into separate breakout rooms for NSs and NNSs to provide tailored instructions. Specifically, NSs received instructions on how to join and use the web interface for the experiment, while NNSs were also taught how to use the AISA. Once NNSs' proficiency with AISA was confirmed, we closed the breakout rooms, shared the web interface entry, and invited participants to join. In the web interface, all participants first completed a practice task to familiarize themselves with the tasks and the system. This was followed by the main tasks, post-surveys, and an interview.

*Tool Introduction.* Prior to commencing the experiment, we introduced NNS participants to AISA using a four-minute tutorial video. This was followed by a brief quiz to ensure that they had understood the basic function of the tool. Once the NNS participants had a thorough grasp of the tool, we encouraged them to engage in simple dialogues with us, in order to verify their ability to use the tool for expressing their ideas.

*Practice Task.* Before the main tasks, each team was asked to work together on the simplified Arctic Survival Task to familiarize themselves with teammates, our system, and the survival task. This practice task consisted of three steps: (1) reading the survival task introduction, for a maximum of two minutes; (2) ranking the survival items provided in order of importance independently, for a maximum of three minutes; and (3) discussing the ranking with teammates, for a maximum of five minutes. During the group discussion, participants were encouraged to convince each other through sound reasoning, expertise, or personal experience.

*Main Task.* Two main survival tasks were conducted: in one, the NNS was allowed to use the AISA; in the other, the AISA was not permitted. The order in which these two tasks were presented to the team was counterbalanced to control for any order effects. Each survival task followed the same three-step procedure of the *Practice Task.* The only difference in the main tasks was the duration of the discussion phase, which was extended to a maximum of ten minutes to allow for more in-depth discussion.

*Post-study Survey.* Upon completion of each main task, we collected NNSs' feedback through a post-study survey, measuring NNSs' self-rated speaking competence, anxiety, and workload during the survival task.

*Interview.* We conducted individual interviews with each NNS via Zoom, each lasting approximately 30 minutes.

### 3.6 Measures

We utilized three types of measurements: system log data, surveys, and semi-structured interviews. To answer RQ1, we measured various aspects of NNSs' inputs, including content, duration, number of modified characters, language preference, and length. For RQ2, we measured NNSs' speaking competence. For RQ3, we measured NNSs' anxiety and workload. Additionally, we used interviews to provide in-depth insights and potential explanations for the observed phenomena.

*3.6.1 Log Data.* We collected the input content of each NNS while using the AISA and categorized it based on content similarities. The identification and merging within each category were performed by two authors of this study. In instances of classification conflicts, the authors engaged in discussions until a consensus was reached. Subsequently, we used the following indicators to examine each input pattern: (1) Input duration: refers to the time span from when an NNS first typed a character in the input box to when they pressed 'enter'. (2) Number of modified characters: refers to the total number of keystrokes made during this input process. For English input, each character addition or deletion was counted as one modification (e.g., when inputting *"this is"*, the user needs to do seven keystrokes). (3) Language preference: refers to the language used by NNSs for input, which can be either exclusively their native language, exclusively English, or a combination of both. (4) Input length: each input string was broken down into a series of tokens, and the number of these tokens was treated as the input length.

*3.6.2 Speaking Competence.* Three items from the scale developed by Duran [22] were used to measure NNSs' speaking competence. We selected the items of grammar, tense, and vocabulary from this scale while removing the item of pronunciation. We removed the pronunciation item because AISA's function was to provide NNSs with speaking references and did not cover pronunciation. Therefore, we did not measure AISA's impact on NNSs' pronunciation. NNSs rated their agreement with statements including *"I had problems with grammar when speaking (R)", "Sometimes I do not use the correct verb tense (R)", "I sometimes misuse words (R)"*, on a seven-point Likert scale ranging from 1 = strongly disagree to 7 = strongly agree (Cronbach's alpha = 0.86, M = 4.27, SD = 1.15). A higher score indicated a higher speaking competence for the NNS.

*3.6.3 Speaking Anxiety.* The Foreign Language State Anxiety Scale [9] was used to measure the anxiety of NNSs during the discussion. NNSs were asked to rate their agreement with statements, including *"I was confident about my ability to communicate with other participants using English. (R)" "I felt relaxed speaking English with other participants. (R)" "I got flustered when other participants communicated things I did not understand." "I felt like I didn't have enough time to think before I had to respond to other participants." "It did not bother me when I did not understand everything other participants were saying. (R)" "I started to panic when I had to speak without preparation." "I was not bothered by other participants communicating quickly. (R)" "I got so nervous that I forget things I know"* on the seven-point Likert scale ranging from 1 = strongly disagree to 7 = strongly agree (Cronbach's alpha = 0.86, M = 3.52, SD = 1.12). A higher score indicated higher anxiety for the NNS.

*3.6.4 Workload.* We measured participants' workload with items from the NASA-TLX scale [19]. Initially, the scale-covered mental demand, temporal demand, effort, frustration, and performance. However, we removed the performance item because of its low reliability relative to the other four. The final items included *"How much mental and perceptual activity was required (e.g., thinking, deciding, calculating, remembering, looking, searching, etc.)?" "How much time pressure did you feel due to the rate or pace at which the tasks or task elements occurred?" "How hard did you have to work (mentally and physically) to accomplish your level of performance?" "How discouraged, irritated, stressed or annoyed did you feel?"* All items were rated on a seven-point Likert scale ranging from 1 = very low to 7 = very high (Cronbach's alpha = 0.85, M = 3.24, SD = 1.03). A higher score indicated a higher workload for the NNS.

*3.6.5 Interviews.* The interview focused on NNSs' usage of the AISA and how it affected their speaking, anxiety, and workload. Questions included general inquiries into their ability to follow and contribute to the discussion (e.g, *"Were you able to follow the discussion in two survival games?"*), and their personal strategies to use the tool, such as whether they preferred to input keywords or full sentences and in which language. Furthermore, interviews delved into comparing their experiences with the AISA to other similar tools they had used, like *"In what ways did you feel that the speech support feature was different from the tools you normally use?"* We also probed the challenges that NNSs faced while using the tool by asking *"Did you encounter any challenge or confusion while using this tool? How did you address them?"* Additionally, NNSs were asked to share the impact of the tool on their speaking, anxiety, and workload, such as *"What do you think of the workload imposed by the tool?"* Finally, NNSs were asked *"How do you think the tool can be improved? Why do you think that would help?"* to provide insights into the future development of AISA.

Each interview was audio-recorded and transcribed with the participants' consent. We used thematic analysis [13] to code the transcript content. This process began with two authors independently coding the transcripts. During data coding, the authors ensured that the themes were connected to the research questions. Following the initial coding, the authors engaged in discussions to understand the rationale behind each code and reached a consensus through a process of comparison and reflection. The authors revisited the transcripts multiple times, ensuring that the codes accurately captured the nuances of the participants' responses. Through this iterative process, the authors were able to discern patterns and formulate overarching themes.

Table 2. Patterns of non-native speakers' input queries while using AISA.

| Input Patterns | Sub-categories | Examples of NNS Input | Corresponding Speaking References |
|---|---|---|---|
| With clear instructions | Seeking word translation | translate yacht into Chinese | The word "yacht" translates into Chinese as "游艇". |
| | Rationalizing decisions | map is most important，给出原因；plastic sheet is important，因为可以帮助我们漂浮在海上[a] | I believe the map of the Atlantic Ocean should be our top priority. With the map, we could determine our location and navigate towards the nearest land. Secondly, I think the 20 square feet of opaque plastic sheeting is crucial. It could potentially assist us in floating on the ocean's surface or serve as protection against the elements. |
| Without clear instructions | Stating viewpoints | 要在ocean上能浮起来[b] | We need to ensure we can stay afloat in the ocean. Survival at sea relies heavily on our ability to remain above water. |
| | Only keywords | rope | The rope could also be useful as a potential life line, especially in a situation where one of us might fall overboard. It would be easier to pull them back in with the aid of a rope. I guess that's another practical use for it. |

[a] English translation: map is most important, give a reason; plastic sheet is important, because it can help us float on the sea
[b] English translation: We need to be able to float on the ocean

## 4 Result

In this section, we first answer RQ1 by identifying two main input patterns among NNSs while using the AISA, which we further categorize into four subcategories. We also present the perceptions of NNSs toward each input pattern, as emerged from the interviews. Following this, we address RQ2 by discussing the impact of AISA on NNSs' speaking. Finally, we respond to RQ3 by introducing how AISA influences NNSs' anxiety and workload.

### 4.1 Input Pattern of AISA (RQ1)

We categorized the input content from NNSs who used AISA in two main types: clear instructions vs. lacking clear instructions. For inputs with clear instructions, NNSs explicitly defined the purpose of the support they hoped to receive from AISA (e.g., *"translate yacht into Chinese"*). Conversely, inputs lacking clear instructions did not specify the NNSs' intentions or needs clearly (e.g., *"ocean, plastic sheet"*). We then further divided these two types into two subcategories each (four categories in total), as shown in Table 2. Specifically, the two subcategories under "clear instructions" are: (1) *seeking word translation* (e.g., *"translate 'yacht' into Chinese"*); (2) *rationalizing decisions* (e.g., *"map is most important, give reasons"*). Under the "lacking clear instructions", the two subcategories are: (3) *stating viewpoints* (e.g., *need float on the ocean*); and (4) *only keywords* (e.g., *rope*). In total, NNSs used AISA 76 times, and the frequency for each category was: five times for *seeking word translation*, 36 times for *rationalizing decision*, 23 times for *stating viewpoints*, and 12 times for *only keywords*.

*4.1.1 Seeking Word Translation.* This refers to the pattern in which the NNS seeks translation for certain words or phrases (e.g., *"translate yacht into Chinese"*). The primary advantage underlying this pattern, as highlighted by NNSs, was the facilitation of comprehension. NNS-5 noted, *"This speaking support feature is very helpful to me. It helps me to understand what others are saying, especially some words that I am not familiar with."* However, there were notable drawbacks underlying this input pattern, particularly in terms of increased effort for NNSs. An example outlined in Table 2 illustrates this issue: when an NNS inputs the command *"translate yacht into Chinese"*, expecting a straightforward translation like '游艇', they instead received a more verbose response: *"The word 'yacht' translates into Chinese as '游艇'"*. This not only required additional input effort but also increased the workload as NNSs had to parse through unnecessary text, contrasting sharply with more direct translation tools that provide concise answers.

*4.1.2 Rationalizing Decisions.* This pattern describes instances where the NNS seeks assistance from AISA to rationalize and articulate their thoughts, such as in requests like *"map is most important, give me a reason."* In these instances, NNSs typically have a preliminary idea in mind—such as the order of items in the survival game—and look to AISA to enhance their argument. When seeking help in this pattern, NNSs found AISA was beneficial in constructing detailed rationales and systematizing the expression. For example, it used sequential markers like *"first,... second,..."* to organize content based on the input from NNSs. NNS-24 commented, *"I would arrange my thoughts and ideas in order of importance and enter them into the tool, which would help me generate a more organized expression, because it is hard for me to come up with better reasons to express them on my own in a short period."*

Four NNSs compared AISA with search engines and discovered that AISA was more proficient in offering contextually appropriate support for articulating their thoughts. Normally, NNSs utilize search engines during dialogues to collect evidence or confirm crucial details. Although search engines supply a lot of information, they do not adapt this information to the user's conversational context, necessitating that users put in extra effort to incorporate it into their dialogue. For example, NNS-30 mentioned, *"Search engines frequently produce vast amounts of vague and irrelevant data, complicating the process of selecting the right information for spontaneous conversations."* NNS-9 emphasized that AISA *"creates a thorough and immediately usable speech script,"* which better aligns with NNSs' perspectives and is more appropriate for the context.

*4.1.3 Stating Viewpoint.* This pattern differs from the previous two patterns, as the NNS did not specify what kind of response they expect AISA to provide, but rather, expressed their own viewpoint in sentences, for example, *"need float on the ocean"*. During the interviews, NNSs indicated that their difficulty in constructing full paragraphs led them to use this pattern, relying on AISA to help articulate their thoughts more completely. For instance, NNS-31 stated, *"Sometimes I'm not sure how to construct a complete sentence based on my opinion."* For this pattern, NNSs appreciated AISA's ability to synthesize their ideas. NNS-25 remarked, *"I like this tool because it can combine some of my scattered viewpoints into a complete sentence, which is much better than Google Translate."*

*4.1.4 Only Keywords.* In this pattern, the NNS only inputs a single word without conveying any intention, for example, entering *"rope"*. The scenario for using this pattern was that some NNSs wanted to use AISA to gain inspiration based on keywords when they found themselves momentarily out of ideas. As noted by NNS-29, *"I would enter some keywords or names of props (e.g., the survival item) into the input box, and then the suggestions generated by the tool would help me generate some new ideas or perspectives."* A further example of this is shown in Table 2, where – when the NNS input *"rope"* – the system combined it with the prevailing conversational context to provide new angles and insights for the NNS: *"The rope could also be useful as a potential lifeline, especially in a situation where one of us might fall overboard. It would be easier to pull them back in with the aid of a rope. I guess that's another practical use for it."*

Besides, some NNSs entered this pattern due to their habitual use of translation tools, leading to a dual experience with AISA. For instance, NNS-24 found AISA more satisfactory than traditional translators, appreciating that AISA provided contextually relevant suggestions that enhanced their conversations. NNS-24 commented, *"I think there is a*

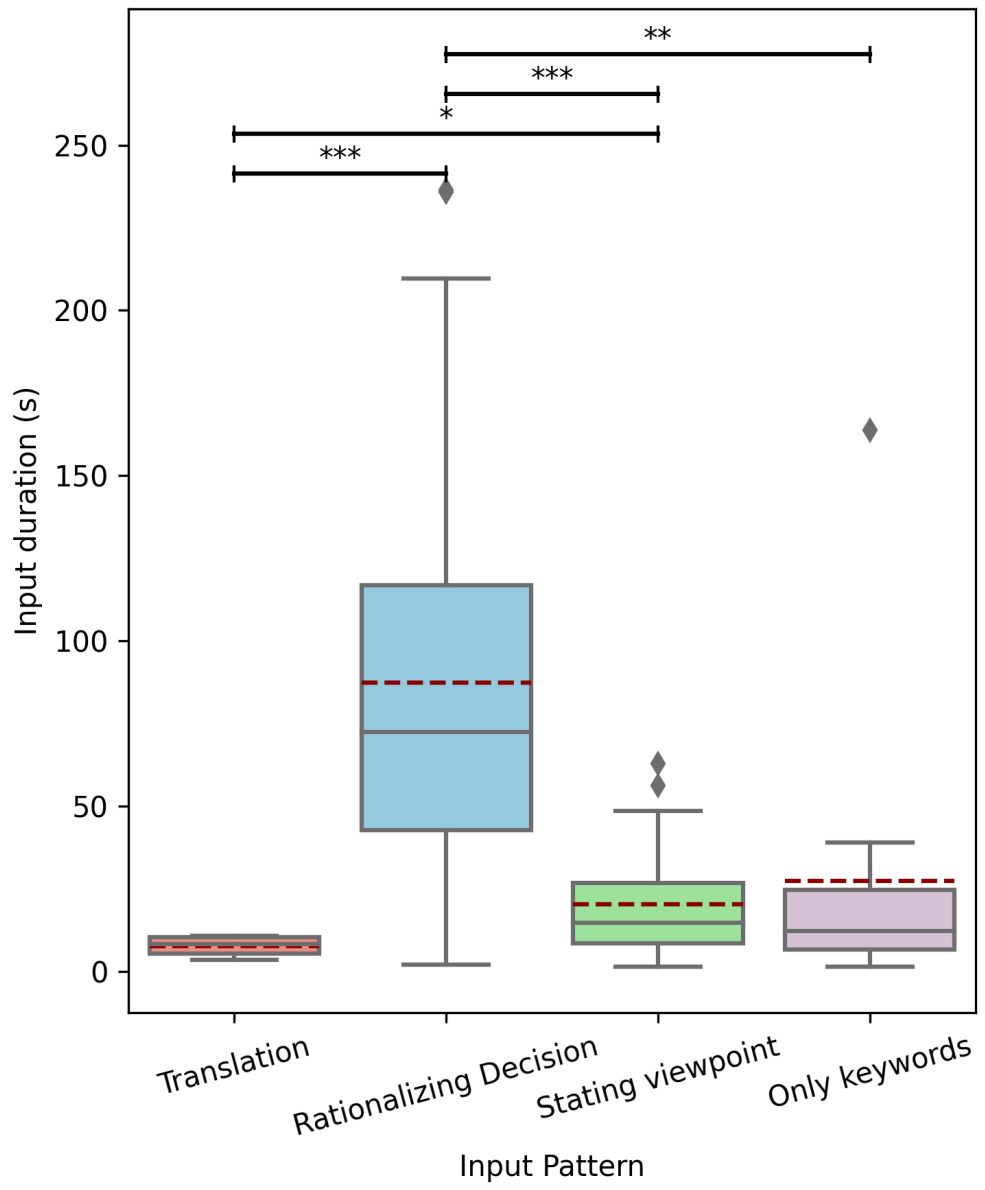

(a) Input duration, by input pattern

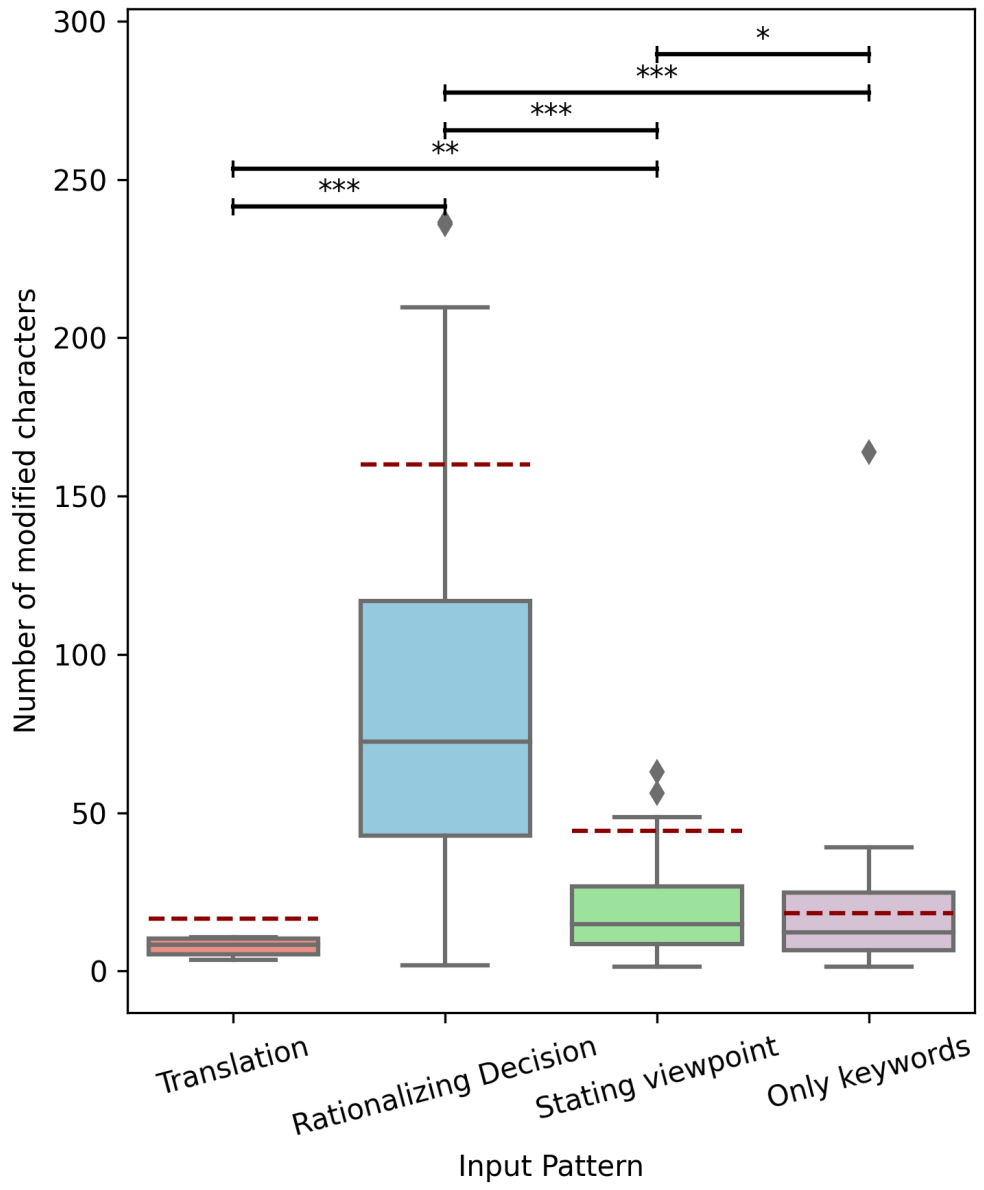

(b) Number of modified characters, by input pattern

Fig. 2. Input duration and the number of modified characters by input pattern. The asterisks denote levels of statistical significance, with * representing p < 0.05, ** representing p < 0.01, and *** representing p < 0.001. The dark red dashed line in the box plot represents the mean of each pattern, and the horizontal bar in the box plot represents median.

*difference between this speech assistant and the simple translation tools that I usually use. The latter may only translate individual words, while this assistant can generate additional content and provide a complete and usable speaking suggestion."* Conversely, some NNSs, accustomed to the standard usage of prevalent translation software, expressed confusion with AISA's different approach. One NNS noted, *"I feel somewhat confused because this differs from the presentation format of previous tools."*

*4.1.5 Input Duration and Number of Modified Characters of Each Input Pattern.* For the four identified patterns, we examined differences in their *input duration* and *the number of modified characters* to provide more insights into the effort expended by NNSs. We first conducted a Levene test for homogeneity of variance and a Shapiro–Wilk test for normality to ensure the data met the assumption for standard parametric analysis. Due to its violation of homogeneity of variance, we performed one-way ANOVA with Brown-Forsythe correction *input duration* (M = 52.3, SD = 59.4) and *number of modified characters* of different patterns (M = 93.2, SD = 107.5). The results revealed significant differences both in *input duration* (F[3, 42.268] = 20.382, p < 0.001) and *number of modified characters* (F[3, 42.856] = 34.908, p < 0.001) among four input patterns, as shown in Fig. 2.

We then conducted the post-hoc analysis using the Games-Howell test. For the *input duration* (Fig. 2(a)), the analyis revealed the significant difference between *rationalizing decisions* (M = 87.260, SD = 65.775) and the other three patterns: *seeking word translation* (M = 7.626, SD = 3.137; p < 0.001, t = 7.206), *stating viewpoints* (M = 20.347, SD = 17.727; p <

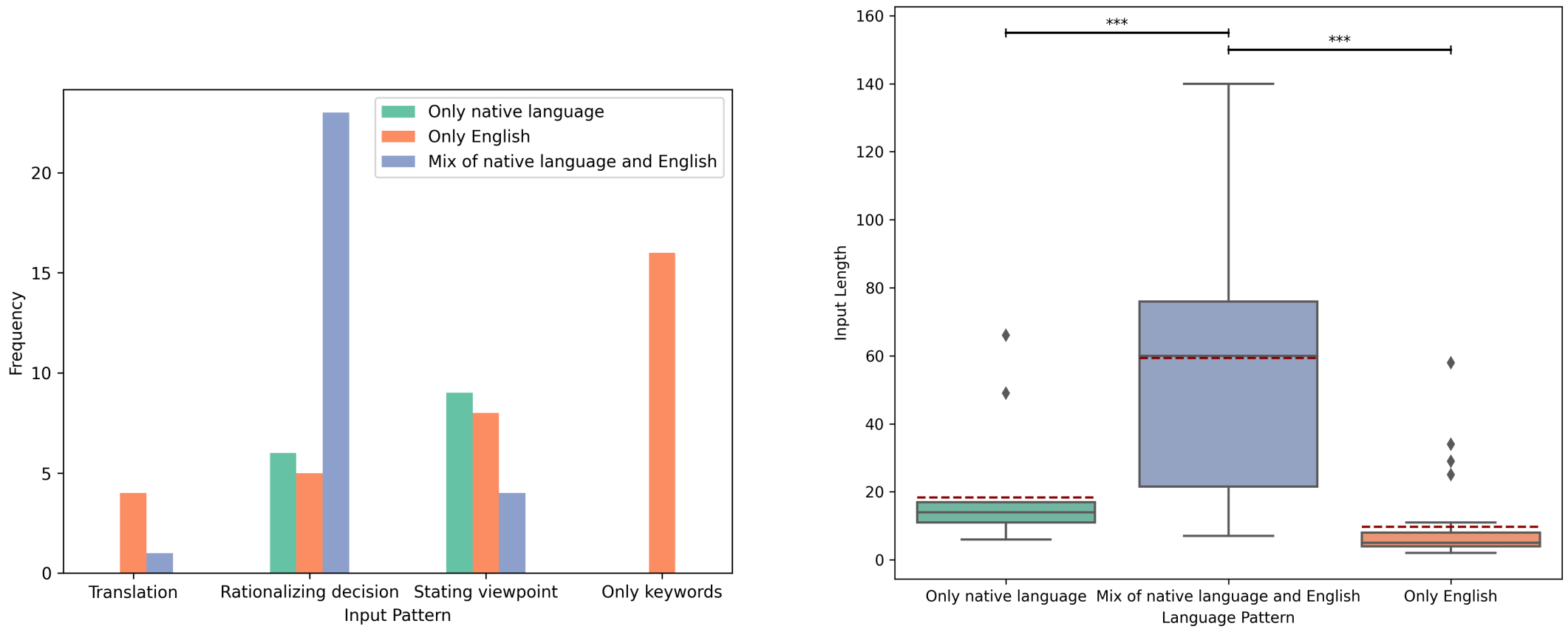

(a) Distribution of the language used across four identified input patterns. Frequency represents the number of input queries.

(b) Input lengths by language preferences. Input length refers to the total number of tokens in each input query. The asterisks denote levels of statistical significance, with * representing p < 0.05, ** representing p < 0.01, and *** representing p < 0.001. The dark red dashed line in the box plot represents the mean of each group, and the horizontal bar in the box plot represents median.

Fig. 3. Distribution of the language used across input patterns and input lengths by each language used.

0.001, t = 5.784), and *only keywords* (M = 27.363, SD = 44.416; $p < 0.01$, t = 3.551). This indicated that the *rationalizing decisions* pattern required the longest input duration among the four patterns.

For the *number of modified characters* (Fig. 2(b)), the analysis also showed that significant differences existed between *rationalizing decisions* (M = 160.000, SD = 123.511) and the other three patterns: *seeking word translation* (M = 16.600, SD = 8.961; $p < 0.001$, t = 6.838), *stating viewpoints* (M = 44.391, SD = 27.853; $p < 0.001$, t = 5.405), and *only keywords* (M = 18.333, SD = 18.676; $p < 0.001$, t = 6.657). This indicated that NNSs made the most modifications to their inputs when attempting to rationalize decisions with AISA.

*4.1.6 Language Preference in Input Patterns.* In their use of AISA, NNSs demonstrated distinct language preferences across different input patterns, including inputs exclusively in English (*Only English*), exclusively in their native language (*Only native language*), and blending both languages (*Mix of native language and English*). The distribution of these language preferences across the four identified input patterns is depicted in Fig. 3(a).

Specifically, for *seeking word translation*, four inputs utilized *Only English*, one instance of the *Mix of native language and English*, and no occurrences of *Only native language*. In *rationalizing decision* pattern, balanced use of *Only English* (n = 5) and *Only native language* (n = 6) were observed, while the *Mix of native language and English* (n = 23) dominated. Regarding *stating viewpoint*, the frequency of using *Only English* (n = 8) or *Only native language* (n = 9) was almost balanced, slightly higher than the frequency of using *Mix of native language and English* (n = 4). The *only keywords* category was characterized by a significant inclination toward *Only English* input, with sixteen instances recorded; meanwhile, there were no instances of *Only native language* or *Mix of native language and English*.

Given that the choice of languages might be influenced by the input length, we further analyzed the link between input length and language preferences to provide greater insights. We first conducted the Levene test for homogeneity of variance and a Shapiro–Wilk test for normality. Given the violation of variance homogeneity, we carried out the one-way ANOVA with Brown-Forsythe correction, which showed that different language preferences displayed a significant difference in input length ($F[2, 45.059] = 29.595$, $p < 0.001$). Post-hoc analysis with Games-Howell, shown in Fig. 3(b), further revealed that the significant differences existed between *Only native language* (M = 18.313, SD = 16.007) and *Mix of native language and English* group (M = 59.323, SD = 42.677; t = -4.743, $p < 0.001$), as well as between the *Only English* (M = 9.655, SD = 12.169) and *Mix of native language and English* group (M = 59.323, SD = 42.677; t = -6.215, $p < 0.001$). The results revealed that when the NNS input the long content, a mixed language was significantly preferred, and when the input content was short, they typically used their native language or English.

Three primary reasons emerged from the interviews explaining the variation in language preference: (1) Simplicity in input. The choice of language was often driven by the desire for simplicity and ease of expression. NNS-20 noted, *"If the English words are too long, I will use Chinese input"*, which highlights a tendency to choose a language that facilitates more straightforward and concise communication. (2) Contextual consistency. Maintaining consistency within the conversational context was a significant factor influencing language choice. For example, NNS-24 explained, *"I wrote it in English because I think the expression in English would be more precise and the translation from Chinese to English could be a bit off."* This preference underscores the importance of using the language employed in the discussion for input to ensure clarity and consistent communication. (3) Logical organization. The ease of organizing thoughts logically in one's native language was also a critical factor, especially when faced with the workload of finding appropriate words in a second language. NNS-13 commented, *"I type in Chinese because I think it is easier for me to express my thoughts in a more logical way."*

### 4.2 The Influence of AISA on NNSs' Speaking (RQ2)

For RQ2, which aims to understand how AISA affects NNSs' speaking, we utilized data gathered from both quantitative measurements and qualitative interviews. To assess the quantitative impact, we measured speaking competence among NNSs both with and without the use of AISA, aiming to detect any significant differences. To examine this, we initially conducted a Levene test for homogeneity of variance and a Shapiro–Wilk test for normality to ensure the data met the assumption for standard parametric analysis. Given the violation of the normality assumption, we opted for the Wilcoxon signed-rank test. The findings indicated no statistically significant differences ($z = 0.171$, $p = 0.875$) in speaking competence between sessions with AISA usage (M = 3.357, SD = 1.204) and without AISA (M = 3.226, SD = 1.254). Despite these results, insights from interviews suggested that AISA was beneficial in aspects not captured by the speaking competence scales that measure grammar, tense, vocabulary, and pronunciation. Particularly, NNSs noted that AISA helped make their speech more on-topic and contributed to the clarity and strength of the arguments. However, NNSs also raised concerns that using AISA diminished their agency, suggesting a complex impact on NNSs' communication.

Specifically, for the benefit of AISA in making NNSs' speech more on-topic, six NNSs highlighted AISA's ability to provide timely content based on their inputted keywords and contextual understanding, which was crucial when their topical knowledge was limited. For example, one common challenge mentioned was not knowing the specific uses of items, such as a nylon rope in the ocean. NNS-26 stated, *"I like that it can generate a complete viewpoint based on the keywords I input."* Similarly, NNS-5 valued the contextual alignment of the content generated by AISA, commenting, *"What I like about this tool is that it can understand our context and all the content it generates is in line with the topic we are discussing. This gives me peace of mind."* Additionally, NNSs reported that AISA was particularly beneficial for enhancing the logicality and the strength of their arguments. Thirteen NNSs noted that when they inputted scattered ideas into AISA, the tool was capable of synthesizing these into a complete sentence or paragraph, imbuing the output with necessary logical structuring to make it appear coherent and organized. Moreover, the strength of the arguments generated by AISA was validated by observing the reactions of NSs. For example, NNS-18 noted, *"There were instances when native speakers expressed appreciation for my speech, especially when I followed the tool's suggestions and provided detailed reasons for my survival-item sequences. At that time, I could see them nodding and choosing to accept my viewpoints."*

Notably, NNSs also expressed concerns regarding AISA potentially depriving them of their agency. Specifically, two NNSs described feelings of being manipulated or controlled by the tool. NNS-18 shared a detailed concern, saying, *"I feel like I might be a little bit manipulated by the system. It's similar to when you provide an answer to a question during an exam without any restriction against looking at the provided answer. Although I know I should come up with a solution on my own, it still tempts me to refer to the answer because I'm aware that it will offer me greater possibilities to get it right."* Further elucidating this point, NNS-5 discussed how reliance on the tool could lead to expressing thoughts that do not genuinely reflect their own, stating, *"If I read it out right after it finishes generating, it may not be what I originally wanted to say; then what I say is not what I think, and I feel like it is controlling me, which I don't like; I think more time is needed for self-review of these contents."*

### 4.3 The Influence of AISA on NNSs' Anxiety and Workload (RQ3)

*4.3.1 Anxiety.* For RQ3, which explores the impact of AISA on NNSs' anxiety and workload during real-time multilingual communication, we analyzed data from both quantitative measures and qualitative interviews. We assessed the levels of anxiety and workload among NNSs when using AISA compared to when they did not use the tool. To check the assumption of quantitative analyses, we first conducted a Levene test for homogeneity of variance and a Shapiro–Wilk test for normality. Since the Shapiro–Wilk test indicated that the data did not satisfy the normality assumption, we employed the Wilcoxon signed-rank test for the analysis. The results indicated no statistically significant difference ($z = 1.201$, $p = 0.235$) between using AISA ($M = 3.712$, $SD = 0.397$) and not using AISA ($M = 3.615$, $SD = 0.325$), demonstrating that there was no significant difference in the levels of speaking anxiety experienced by NNSs whether they used AISA or not.

However, during the interviews, NNSs recognized several factors associated with the use of AISA that might increase their anxiety. Notably, the anxiety discussed was not specifically about speaking; instead, it was a broader sense of anxiety stemming from the additional tasks required by AISA, such as inputting and reviewing the output, which not only conflicted with the urgency of real-time communication but also added to the already high workload of the NNSs. NNS-27 highlighted these challenges by stating, *"It increases my anxiety because I need extra time to input keywords, wait for feedback from the tool, and then convert the feedback into content that I can speak out loud."* NNSs further expressed concerns about the AISA's accuracy in reflecting their intended meaning, adding to their apprehension. For example, NNS-15 remarked, *"Before I express myself, this tool makes me feel a little anxious because I worry about whether the content it generates can accurately convey my meaning."* Additionally, keeping up with the pace of conversations while using AISA proved to be stressful, given the already high-workload situation of NNSs. NNS-27 explained, *"I sometimes feel anxious because I need to use this tool while following the conversation. Sometimes, the speed of content generation by the tool cannot keep up with the pace of the conversation, causing me to miss some parts of the conversation."*

*4.3.2 Workload.* Similar to the examination of anxiety, we first examined the assumption of equal variance and homogeneity and the result of the Shapiro–Wilk test showed that our data did not satisfy the normality assumption. Therefore we conducted a Wilcoxon signed-rank test, which revealed no statistically significant difference ($z = 0.882$, p

= 0.383) between not using AISA (M = 3.313, SD = 0.917) and using AISA (M = 3.438, SD = 0.980). This indicated that using AISA did not significantly affect the workload of NNSs compared to not using the tool.

During the interviews, NNSs conveyed mixed feelings about the workload associated with using AISA. While the necessity of multitasking with AISA increased their workload, the tool also helped reduce their workload by providing additional knowledge and logically structured references.

Specifically, for the increasing workload led by AISA, NNSs mentioned that in addition to the usual demands of listening, processing, and speaking during multilingual communication, they also had to manage additional tasks introduced by AISA, which included (1) inputting information into the system, (2) reviewing system-generated responses, and (3) integrating these responses with their personal ideas. Twenty NNSs described the task of inputting information into the system as particularly challenging. NNS-7 noted, *"the workload of this process is higher"* compared to engaging in unassisted conversation. This challenge was exacerbated for some by their slower typing speeds, which made it difficult to switch tasks effectively during conversations. NNS-6 expressed frustration with her inability to manage these demands due to her slow typing speed, *"For a slow typist like me, it can be a struggle at times because I have to juggle between listening to what the other participants are saying and typing the keywords at the same time, and sometimes I feel distracted, so I didn't use the tool after trying it twice."* Furthermore, the complexity of the responses generated by AISA sometimes added to NNSs' cognitive strain. Four NNSs reported that the system occasionally used unfamiliar words or complex syntax, which increased their workload. NNS-29 remarked, *"Sometimes the tool generates suggestions that I don't fully comprehend, such as words I have never seen before."* Additionally, three NNSs found it overwhelming to combine system-suggested content with their own ideas. NNS-18 explained, *"I have to think through the answers in my head and then select some of them and integrate them into my expression, which takes some time and effort."*

In the meanwhile, NNSs also noted a decrease in workload due to AISA's help in (1) offering insights that aided in comprehending conversational contexts and forming complete speech, and (2) simplifying the process of formulating thoughts. Specifically, AISA equipped NNSs with rapid access to essential knowledge, enabling them to assimilate new concepts into their existing understanding and thereby sparking insights. For instance, NNS-11 was initially puzzled upon encountering the item *"plastic sheeting"* during the Ocean survival task. AISA provided a suggestion that he could use plastic bags to collect rainwater or evaporated seawater, clarifying the concept and its application. NNS-11 explained, *"This gave me a sudden realization and allowed me to have a greater sense of immersion, deepening my understanding of this scene. It also made me think about other items in the same way."* Additionally, by providing convenient expansions and rationalizations, the system streamlined the process for NNSs to develop their initial fragmented thoughts into complete ideas, reducing their effort to do so on their own. NNS-3 stated, *"When there are tools, I can express my opinions more fully. But in the game without the tool, I have some ideas but how to say them specifically is still under consideration, so they are not expressed very clearly."*

## 5 Discussion

We discovered four input patterns of using AISA among NNSs during real-time multilingual communication, each reflecting different assistance needs. Specifically, the pattern *rationalizing decision* posed a significant burden on NNSs, as evidenced by longer input times and more modifications to the input. Although the quantitative data showed that AISA did not significantly alter the speaking competence of NNSs, as an exploratory study, the qualitative interviews provided valuable insights into the user experience with the tool. Specifically, they uncovered improvements in areas not captured by quantitative measurements, such as the logical flow, relevance, and richness of speech. Additionally, these interviews also indicated that AISA potentially increased NNSs' anxiety due to the additional tasks it required, such as inputting queries and reviewing responses. It also had mixed effects on NNSs' workload, which may help explain why speaking competence did not improve. These findings also provide empirical insights for the AIMC field, highlighting the importance of carefully addressing the risks posed by workload and anxiety, which can counteract the intended effects of synchronous assistance tools during real-time use.

In the subsequent discussion, we first discuss the effort expended by NNSs in acquiring AISA support and potential strategies to mitigate this effort, based on the identified input patterns when exploring RQ1. We then discuss potential reasons that may hinder the enhancement of NNSs' speaking competence and provide insights into future design aimed at enhancing NNSs' speaking and reducing their anxiety and workload, as derived from the findings related to RQ2 and RQ3.

### 5.1 Minimizing NNSs' Efforts in Requesting Help from the Speaking Assistance

As an initial attempt to support NNSs' speaking in real-time multilingual communication, our study observed that NNSs exerted considerable effort when requesting the tool to help them rationalize decisions, and it proves crucial to support NNSs in using their native languages when using the speaking assistance tool during real-time multilingual communication. Based on these insights, we suggest strategies that streamline the process of rationalizing decisions and constructing inputs that accommodate NNSs' use of native languages.

Specifically, our research found the pattern *rationalizing decision* required the most effort, as evidenced by the longest input duration and the highest number of character modifications. This increased effort can be attributed to the two-step process NNSs must engage in when using the system to help rationalize decisions: (1) formulating a rough idea in their mind, such as *Importance: torch > sunglass > First-aid kit* and (2) compressing these ideas into the shortest possible format for easier input. Compared to other usage patterns such as simply asking for translations, these steps inevitably lead to increased cognitive processing [32, 56]. Particularly in dynamic settings of real-time multilingual communication, where NNSs' ideas continuously evolve based on others' contributions [6, 17], NNSs might need to modify inputs frequently. For example, an NNS inputs an initial idea and requests the system to rationalize it; however, as new ideas are introduced by NSs during the discussion, the NNS may be influenced and feel the need to adjust their input. This dynamic interaction necessitates additional time and revisions when asking the system to help rationalize decisions.

Additionally, our study highlights the critical importance of supporting NNSs in using their native languages when using the speaking assistance tool. We observed that NNSs frequently employed a combination of a common language and their native language to construct long and logically complex inputs, a multilingual approach that benefits ease of input and enhances the logical organization of thoughts. Particularly, when NNSs used a mixed language for input, especially in the *rationalizing decisions* mode, it could reduce the cognitive burden of finding appropriate vocabulary, thus preserving cognitive resources for better expressing their idea and logical reasoning. In line with these observations, previous research has also emphasized the vital role of the native language in second language acquisition, highlighting its contribution to cognitive scaffolding, reducing learner anxiety, promoting deep understanding, and supporting positive cross-linguistic transfer [20, 28, 82].

Taken together, these insights highlight the necessity of designing speaking assistants that allow NNSs to take advantage of their native language for support, which may also reduce the effort NNSs must expend in seeking help to rationalize decisions. A straightforward approach is to enable NNSs to use their native language vocally during conversations as input queries. This is because expressing ideas in one's native language can facilitate deeper thinking, richer content, and quicker construction. Moreover, voice input speeds in English are 3.0 times faster compared to typing, and in Mandarin, the speeds are 2.8 times faster [73]. Therefore, integrating voice input could be a potential solution to minimize the effort caused by frequent manual input, which poses a major burden for NNSs during real-time communication.

### 5.2 Supporting NNSs' Speaking While Considering Personalization, Autonomy, and Trust

Our study revealed that while quantitative data indicated no significant impact of AISA on NNSs' speaking competence, qualitative interviews highlighted critical aspects not captured by these measurements, notably that AISA made NNSs' speech more on-topic and enhanced both the clarity and the strength of their arguments.

However, it is also noted that AISA raises several concerns including insufficient personalization and diminishing the autonomy of NNSs. Moreover, in the long run, hallucination of LLM can also lead to trust issues, limiting its practical application. The subsequent sections delve deeply into these concerns, offering recommendations on how future endeavors should focus on enhancing design to prevent such issues.

*5.2.1 Matching Suggestions to NNSs' Proficiency and Expression Styles.* Regarding the non-significant results in NNSs' speaking competence, one possible explanation is that NNSs may struggle to comprehend complex words within AISA's suggestions and often need to adjust them, which introduces new errors in grammar, vocabulary, or tense due to their limited English proficiency (see Section 4.3). Another possible reason why AISA may have limited benefits for some NNSs in speaking competence is the mismatch between the system's suggestions and the users' personal expression styles. Due to individual differences in syntax and communication preferences, the generic suggestions often fail to resonate with NNSs, leading them to rely on their own expressions instead.

To address these issues, future design can employ retrieval augmented generation (RAG) system [99] for the personalization of AISA's output to align more closely with individual language proficiency and linguistic habits since people are more likely to satisfy personalized content [81]. Specifically, it can interface with users' existing personalized data (such as vocabulary learning and speaking practice records) and their ongoing conversation context to generate suggestions that reflect the user's habitual word choices, syntactic patterns, and stylistic preferences, which can help create a seamless and intuitive experience, reinforcing the tool's usability and effectiveness.

*5.2.2 Supporting NNSs Without Compromising Autonomy.* Despite the lack of significant impact on NNSs' speaking competence, interviews frequently highlighted that AISA made speech more on-topic and enhanced both the clarity and the strength of the arguments (see Section 4.2). This underscores AISA's unique benefits over other language support tools, such as translation tools, demonstrating its capacity to address deeper communicative needs beyond mere linguistic accuracy. Overall, the challenges NNSs face in speaking during real-time multilingual communication can be categorized into two levels: (1) linguistic issues [42, 94, 95], such as problems with vocabulary, grammar, and tense; and (2) communication issues, such as not knowing how to organize and express ideas more clearly or how to make arguments stronger [75]. Interviews indicated that the more significant advantage of AIMC in speaking lies in providing support for the second category.

Meanwhile, agency concerns have been voiced by certain NNSs. They felt overshadowed by the suggestions provided by AISA, merely "reading out" the content, which led to feelings of domination, idea fixation, and frustration. This issue is critical as it could further limit user creativity and hinder their ability to explore solutions independently [37, 48, 69, 92]. Moreover, this issue reflects a broader phenomenon observed in AI ethics, where there is an increasing demand for greater agency in using AI technologies. Users emphasize the need to act independently and make their own choices rather than feeling coerced by external systems [16, 40, 57]. The field is thus emphasizing the importance of balancing automated assistance with human agency [54]. In response to these concerns, especially in the context of supporting speaking in real-time multilingual communication, it is vital to ensure that the references provided by speaking assistance tools do not overly constrain users' ideas or dominate their decision-making process. One potential solution is to avoid offering complete sentences and instead provide discrete elements such as words, phrases, and short sentences. This is because providing improper examples can suppress ideas in creative scenarios [61, 80], while providing discrete elements ensures that users have more space for imagination [55, 96], therefore potentially enhancing novelty and autonomy of NNSs.

*5.2.3 Addressing Potential Hallucination Issues.* In this study, although AISA was not reported to produce hallucinations, it is still a prevalent challenge in LLM applications [36]. Therefore, as AISA is applied more broadly, the occurrence of hallucination issues cannot be entirely ruled out. Particularly, this poses a potential risk for NNSs with weaker language skills, as they may not recognize hallucinated outputs and might directly use the suggestions provided by the LLM, leading to potential embarrassment. Moreover, when NNSs do recognize hallucinations, it could undermine their trust in the tool, as previous research shows that the accuracy of LLM-generated information is a critical factor influencing user attitudes and trust toward LLM-based communication tools [2, 51, 62].

To address hallucination issues, one potential solution is leveraging uncertainty visualizations or confidence indicators for the generated suggestions, which can enhance the transparency of AI-supported interactions [62, 76]. Such visual cues would allow users to assess the reliability of outputs at a glance, fostering trust and enabling more informed decision-making during real-time interactions. Another possible approach is implementing a RAG system [11, 79, 99]. Specifically, RAG frameworks first retrieve relevant information from a trusted and dynamically updated external knowledge base (e.g., domain-specific documents, curated datasets, or real-time sources) based on the input query. This information is then integrated into the LLM's generative process as additional context, ensuring that the outputs are grounded in accurate and verified data. By constraining the generation process with reliable information, RAG could reduce hallucinated responses and improve NNSs' trust in AISA across various scenarios.

### 5.3 Alleviating NNSs' Anxiety and Workload When Using the Supporting Tool

Our findings highlighted that introducing support tools for NNSs in real-time multilingual communication could potentially deepen their anxiety and increase their workload. This effect arises because the context demands timely interaction and requires NNSs to engage multiple cognitive processes simultaneously, such as understanding others' speech while formulating their own viewpoints [45, 67, 84]. Utilizing such tools introduces additional tasks, compounding the complexity of multitasking. In our context, extra tasks primarily involve entering information and reviewing the results.

Consequently, in real-time multilingual communication that demands quick exchanges, the need to perform additional tasks with the tool can make NNSs feel pressured or embarrassed, as they require moments to operate the tool, potentially causing others to pause and wait, leading to awkward silences and communication breakdowns. To address this issue, one of the potential solutions is to integrate an intelligent agent into discussions. Such an agent could actively encourage NNS participation and suggest giving NNSs time to think. Previous research has shown that including an agent intervention in multilingual discussions can lead to more balanced participation [45] and that humans tend to perceive artificial intelligence as a social actor and will attribute responsibility to the AI [18]. Therefore, in such settings, having the AI agent initiate actions for NNSs can help both NNSs and NSs attribute the action to the AI, not the NNSs, reducing the NNSs' anxiety about taking time to use the tool and the anxiety associated with potential communication breakdowns.

Another source of anxiety and workload for NNSs stems from uncertainty about the system-generated suggestions, compounded by a general fear of making mistakes or performing poorly [15, 35, 66]. One contributing factor to such uncertainty is the delay in suggestion generation, where responses are only displayed after being fully generated. This waiting period, coupled with doubts about whether the suggestions will adequately meet their needs, may heighten NNSs' anxiety. Additionally, if the suggestions are unsatisfactory, NNSs face further stress from having to modify the prompt and endure another waiting period. To alleviate these issues, future systems could adopt a "streaming mode" for generating suggestions that display them incrementally, either letter by letter or in segmented blocks. This method would ensure that NNSs receive partial content more promptly and reduce the anxiety associated with waiting for complete suggestions, as NNSs can begin to evaluate and use parts of the suggestions as they appear.

Additionally, future research could consider promoting mutual understanding between NNSs and NSs to alleviate the anxiety and workload NNSs experience during real-time communication. As found in prior research, NSs often fail to share the responsibility of communication [30, 47, 74], leaving the burden disproportionately on NNSs. To address this imbalance, systems could encourage NSs to take on part of the responsibility. For instance, a bidirectional communication channel could be developed: when NNSs seek assistance using a tool, the system could automatically send a signal to NSs, indicating that the NNS is facing difficulties and needs more understanding and support, as proactive self-disclosure about challenges can promote prosocial behaviors and enhance empathy from the helper [89, 98]. With greater understanding and patience from NSs regarding the time NNSs need to use the tool, NNSs may experience reduced anxiety and less urgency to produce immediate responses, thereby decreasing their workload.

### 5.4 Implications for Real-Time AI-Mediated Spoken Communication

AIMC has frequently been applied to help people overcome linguistic barriers, such as improving text-based communication, fostering idea generation, and providing emotional and contextual support [23, 24, 31, 49, 77]. However, AI-mediated real-time spoken communication remains a relatively underexplored area. By exploring AI-mediated spoken communication, our study offers valuable insights for future research in this domain. On one hand, we demonstrate that AISA can enhance speaking content, which aligns with prior studies showing that AIMC can improve the quality of communication content [10, 23, 24, 34, 46]. On the other hand, our findings highlight the critical importance of addressing anxiety and workload in AI-mediated spoken communication - a consideration often overlooked in the design of current asynchronous AIMC tools. Specifically, spoken communication in synchronous settings presents distinct challenges, such as increased cognitive multitasking and the need to manage immediate conversational dynamics. Hence introducing tools in such settings may add workload, potentially undermining the tools' benefits. When developing tools for AI-mediated spoken communication, future research should consider these factors to ensure that the tools effectively support communication without imposing further cognitive burdens on users.

Additionally, our study suggests that future research on AI-mediated spoken communication could expand evaluation metrics to include aspects beyond speaking competence, particularly to measure the quality of users' communication content. This recommendation stems from our qualitative interviews, which revealed that AISA helped NNSs improve the logical flow, on-topic nature, and richness of their speech factors not captured by our quantitative measurements. Therefore, in order to provide a more comprehensive evaluation of the tool's impact in AI-mediated spoken communication, future research could include scales like those measuring users' communication self-efficacy [68] and perceived communication quality [91].

## 6 Limitation

This study has several limitations. First, the primary language used was English, potentially limiting the generalizability of our findings to other linguistic contexts. Additionally, the NNSs participating in the study were predominantly studying in English-speaking countries, which suggests that their English language skills were relatively adequate. Furthermore, individuals from different cultural backgrounds may interact with AISA in varying ways, as prior work shows that culture influences technology acceptance and usage [45, 53, 93]. For example, Japanese users, who are often highly concerned about disturbing others [45], may find it more challenging to use AISA because it might require NSs to wait while Japanese NNSs use the tool. To address these limitations, future research should explore scenarios where the common language used extends beyond English, include NNSs with lower levels of English proficiency, and involve NNSs from diverse cultural backgrounds beyond Chinese.

Second, we chose survival tasks as our experimental setting, following previous research on NNS and NS collaboration [44], but these tasks may not reflect scenarios where personal benefits, like teamwork in companies, are at stake. Consequently, participants might be more inclined to conform or accept others' ideas without extensive communication, possibly limiting the interactions between NNS and NS. Future research should consider linking participant performance with tangible rewards or select tasks that necessitate more intensive conversation, such as debates, to encourage richer communication.

Third, we did not explore the effects of AISA on NSs or how NSs perceive AISA. We made this decision in this study because the tool is primarily designed to serve NNSs, and our main concern was its impact on them, but it is important to recognize that NSs might also experience changes due to the tool's use by NNSs or notice differences in overall communication dynamics. Future research should include examinations of NSs' experiences to provide a more comprehensive insight into the effects of such tools.

Fourth, we did not measure participants' prior experience with LLMs, which could have influenced the outcomes [97]. Although we provided comprehensive training to familiarize participants with the tool, their previous experience with LLMs might still have affected their interactions with AISA. Future research could assess participants' prior LLM experience to gain a deeper understanding of how such tools are used during AI-mediated spoken communication. Additionally, while we did not observe hallucination issues in our study, future research could conduct long-term field studies to examine potential hallucinations of AISA and their impact on NNSs, particularly regarding NNSs' trust in the tool.

Lastly, since AISA primarily provides speaking references for NNSs, it does not target pronunciation. However, pronunciation remains a significant speaking challenge for NNSs. Future research could explore ways to address this issue. Moreover, since our quantitative data showed that AISA did not significantly improve NNS speaking competence, and our qualitative data suggested that AISA might increase cognitive load, future research could explore ways to enhance speaking competence while reducing cognitive load, such as personalizing tool features for individual users to minimize user effort in input and make them more comfortable using the generated content.

## 7 Conclusion

In this study, we introduced the AI-based Speaking Assistant (AISA), which leverages Large Language Models to enhance NNSs' speaking in real-time multilingual communication. Our research identified four distinct input patterns of AISA among NNSs, highlighting diverse needs, with the pattern "rationalizing decisions" being particularly effort-intensive. Despite quantitative data not showing significant changes in NNSs' speaking competence, qualitative insights revealed

AISA's role in enhancing both the clarity and the strength of NNSs' arguments, aspects not captured by quantitative measures. Interviews also indicated that in the context of real-time multilingual communication, where there is inherent urgency and a high workload on NNSs, the multitasking required by the assistance tool can potentially further increase NNSs' anxiety and workload. Based on these insights, we propose design implications and hope future tools will better support NNSs in actively participating in multilingual communication.

## 8 Acknowledgement